\documentstyle[prl,aps]{revtex}
\begin{document}
\draft

\title{On the Cosmic Rotation Axis}

\author{Rainer W. K\"uhne}

\address{Fachbereich Physik, Universit\"at Wuppertal, 42097 Wuppertal, 
Germany, kuehne@theorie.physik.uni-wuppertal.de}

\maketitle

\begin{abstract}
Recently, Nodland and Ralston reported to have discovered a cosmic 
axis. We argue that their axis is supported by an earlier 
independent observation on the spin axes of galaxies in the 
Perseus-Pisces supercluster and explainable within the framework of 
G\"odel's cosmology.
\end{abstract}


Nodland and Ralston \cite{one} claim to have found evidence that polarized 
radiation emitted by distant radio galaxies shows residual rotation, 
remaining after Faraday rotation is extracted, that follows a 
dipole rule. The residual angle $\beta$ appears to be correlated with 
the distance $r$ of the radio galaxy and the angle between the 
propagation wave vector ${\bf k}$ of the radiation and the axis ${\bf s}$. 
In sum, the claimed relation for the mean residual angle is
\begin{displaymath}
< \beta > = \frac{r}{2\Lambda_{s}} \cos ({\bf k},{\bf s}).
\end{displaymath}
Expressed in the coordinates $R.A.$ = right ascension and $decl.$ = 
declination the axis points in the direction
\begin{displaymath}
{\bf s} = (R. A. ,decl.) = (315^{\circ}\pm 30^{\circ}, 0^{\circ}\pm 
20^{\circ}).
\end{displaymath}
Under the assumption of the Einstein-de Sitter cosmology (i. e. critical 
mean mass density and vanishing cosmological constant) and by setting the 
present value of the Hubble parameter to
\begin{displaymath}
H_{0}=(1.5\times 10^{10}\mbox{yr})^{-1}=66.7 \mbox{km/(s~Mpc)},
\end{displaymath}
the birefringence scale was found to be
\begin{displaymath}
\Lambda_{s}=(1.1\pm 0.08)\times 10^{25}\mbox{m}.
\end{displaymath}
The corresponding universal angular velocity is
\begin{displaymath}
\omega = \frac{c}{2\Lambda_{s}} = 1.4\times 10^{-17}\mbox{s}^{-1}.
\end{displaymath}
Such an intrinsic spin would have dramatic consequences, because it would 
violate Mach's principle which states that, ``A system on which no forces 
act is either at rest or in uniform motion relative to the fixed stars 
idealized as a rigid system''.

A cosmology with intrinsic universal rotation was already suggested by 
G\"odel \cite{two}. Although its original version considers the static 
universe 
only, this model becomes viable if universal expansion is included, as shown 
by Raychaudhuri \cite{three}.

G\"odel predicted that the original order of the rotation axes of 
galaxies has been parallel to the universal rotation axis. This idea 
was supported by the discovery \cite{four} that the distribution of the 
rotation 
axes for both the spiral and ellipsoid galaxies of the filament-like 
Perseus-Pisces supercluster is bimodal. One of the peaks is roughly aligned 
with the major axis of the supercluster while the second peak is roughly 
$90^{\circ}$ from the first. Quite remarkably, the major axis of this 
supercluster points in the direction
\begin{displaymath}
{\bf s}_{pp}=(R.A.,decl.)
            =(290^{\circ}\pm 20^{\circ},-20^{\circ}\pm 10^{\circ}).
\end{displaymath}
Within the error bars this axis agrees with the one detected by Nodland 
and Ralston.

\end{document}